\documentclass[conference]{_sty/IEEEtran}
\IEEEoverridecommandlockouts
\usepackage{cite}
\usepackage{caption}
\usepackage{float}
\usepackage{stfloats}
\usepackage{amsmath,amssymb,amsfonts}
\usepackage{threeparttable}
\usepackage{graphicx}
\usepackage{textcomp}
\usepackage{xcolor}
\usepackage{algorithm}
\usepackage{algpseudocode}
\usepackage{multirow}
\usepackage{multicol}
\usepackage{float}
\newcommand{\tabincell}[2]{\begin{tabular}{@{}#1@{}}#2\end{tabular}}
\def\BibTeX{{\rm B\kern-.05em{\sc i\kern-.025em b}\kern-.08em
    T\kern-.1667em\lower.7ex\hbox{E}\kern-.125emX}}
\begin{document}

\title{A Collaborative PIM Computing Optimization Framework for
Multi-Tenant DNN

}

 \author{\IEEEauthorblockN{Bojing Li}
 \IEEEauthorblockA{\textit{Computer Science and Electrical Engineering Department} \\
 \textit{University of Maryland, Baltimore County}\\
 Baltimore, USA \\
 ji18978@umbc.edu}
 \and
 \IEEEauthorblockN{Duo Zhong}
 \IEEEauthorblockA{\textit{Computer Science and Electrical Engineering Department} \\
 \textit{University of Maryland, Baltimore County}\\
 Baltimore, USA \\
 duoz1@umbc.edu}
 \and
 \IEEEauthorblockN{Xiang Chen}
 \IEEEauthorblockA{\textit{Department of Electrical and Computer Engineering} \\
 \textit{George Mason University}\\
 Fairfax, USA \\
 xchen26@gmu.edu}
 \and
 \IEEEauthorblockN{Chenchen Liu}
 \IEEEauthorblockA{\textit{Computer Science and Electrical Engineering Department} \\
 \textit{University of Maryland, Baltimore County}\\
 Baltimore, USA \\
 ccliu@umbc.edu}
 }

\maketitle

\renewcommand{\thefootnote}{}
\footnotetext{This work is partially supported by NSF CNS-2239638.}

\begin{abstract}

Modern Artificial Intelligence (AI) applications are increasingly utilizing multi-tenant deep neural networks (DNNs), which lead to a significant rise in computing complexity and the need for computing parallelism.
ReRAM-based processing-in-memory (PIM) computing, with its high density and low power consumption characteristics, holds promising potential for supporting the deployment of multi-tenant DNNs. 
However, direct deployment of complex multi-tenant DNNs on exsiting ReRAM-based PIM designs poses challenges. 
Resource contention among different tenants can result in sever under-utilization of on-chip computing resources. 
Moreover, area-intensive operators and computation-intensive operators require excessively large on-chip areas and long processing times, leading to high overall latency during parallel computing.
To address these challenges, we propose a novel ReRAM-based in-memory computing framework that enables efficient deployment of multi-tenant DNNs on ReRAM-based PIM designs. 
Our approach tackles the resource contention problems by iteratively partitioning the PIM hardware at tenant level.
In addition, we construct a fine-grained reconstructed processing pipeline at the operator level to handle area-intensive operators. 
Compared to the direct deployments on traditional ReRAM-based PIM designs, our proposed PIM computing framework achieves significant improvements in speed (ranges from 1.75$\times$ to 60.43$\times$) and energy(up to 1.89$\times$). 

\end{abstract}


\section{\textbf{Introduction}}

Processing-in-memory computing enables specific computations to be performed in memory, which reduces power consumption and processing latency by minimizing the overhead of data transmission between memory and processors~\cite{chen2016eyeriss}. 
One of the most promissing PIM techniques is the resistive random-access memory~(ReRAM)-based PIM design. 
The resistive devices, a.k.a., memristors can represent data by changing their internal resistance level and the formed crossbar array structures can perform storage and matrix calculation. 
A variety of ReRAM-based PIM designs have been proposed for DNN computing~\cite{chen2018regan,zheng2020lattice,mao2018versatile}.
The proposed designs can support one DNN or several different DNNs executions with certain reconfigurability, while only a single DNN is supported in one execution cycle.

However, with the rapid development of AI applications, AI computing, particularly the computing of deep neural networks (DNNs), has evolved into multi-tenant scenarios~\cite{jeon2019analysis,yu2022survey}. 
In these scenarios, multiple DNNs are processed simultaneously, requiring parallel deployment on computing platforms. 
This approach results in high transmission overhead due to the computation of parameters from multiple networks. 
Additionally, the results from individual networks in multi-tenant DNNs are often interrelated, making the overall parallelism and latency of each tenant crucial.
The multi-tenant DNNs were observed and their processing was investigated for several specific applications, such as autopilot~\cite{huang2021close} and robot moving~\cite{jeon2016multi}.
Although ReRAM-based PIM designs have been explored for several advanced deep learning algorithms, such as transfer learning~\cite{chen2018emat}, natural language processing~\cite{tambe2021edgebert}, etc., critical challenges when deploying multi-tenant DNNs on these PIMs still exist.


Traditional ReRAM-based PIM architectures usually have multiple unified processing cores (or elements), each of which has numerous ReRAM crossbar arrays. 
In such architecture, hardware resource will become severe insufficient when multiple large-scale tenants are deployed simultenously.
Therefore, the multiple tenants need to be depolyed sequentially in different clock cycles.
However, resource under-utilization will occur easily due to varied computing resource required by different tenants.
Simply dividing hardware cores into multiple regions for each tenants can enable parallel processing, while the resource under-utilziation still remains unsolved and overall computing latency is constrained by the slowest tenant.
As such, a hardware and tenant aware resource allocation strategy is necessary to improve the hardware utiliztion and hence computing speed.

\begin{figure}[t]
  \setlength{\abovecaptionskip}{1.4mm}
  \setlength{\belowcaptionskip}{-0.cm}
  \centering
  \includegraphics{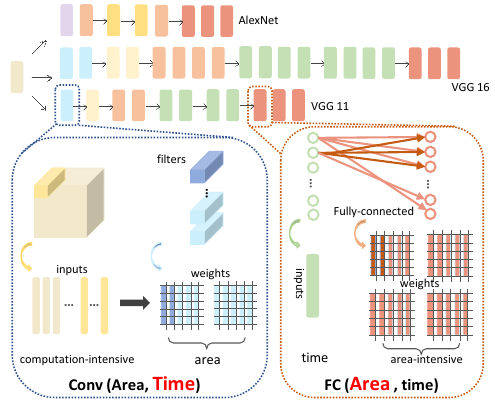}
  \caption{Single DNN Deployment on ReRAM-based PIM Accelerators}
  \label{fig1}
  \vspace{-8mm}
\end{figure}


In addition, as is shown in Figure~\ref{fig1}, different types of operators within a tenant have highly different hardware resource needs when deploying on the ReRAM-based PIM designs. 
For example, area-intensive operators such as 
fully connected layers~(fc) and computation-intensive operators such as convolutional layers~(conv) have high area/computation requirements respectively. 
Area-intensive operators usually require a high volume of on-chip computing areas, leading to the stagnation of other processing pipelines. 
On the other hand, computation-intensive operators usually requires a heavy amount of processing cycles and slow down overall latency. 
Previous research adopts slicing~\cite{ankit2020panther} to split operators but it is insufficient in the parallelized multi-tenant scenarios. 
There are also recent research in splitting and rescheduling operators to overcome the resource under- and over-utilization problems and improve the computing efficiency in the multi-tenant DNN computing on GPU. 
However, a fine-grained multi-tenant DNN computing pipelines to reduce the overall latency is still unexplored in the ReRAM-based PIM designs.

Therefore, we propose a joint optimization framework with multi-tenant DNN-aware hardware resource allocation and a fine-grained operator processing pipeline to address these challenges. The tenant-level hardware resource allocation and operator-level pipeline reconstruction interact dynamically: the hardware allocated to a single DNN (or tenant) affects the granularity of operator splitting, and the processing efficiency of the fine-grained pipeline post-reconstruction determines the hardware resource demands.
This joint optimization framework significantly improves resource utilization and reduces computing latency. We employ a classic ReRAM-based PIM design (i.e., ISAAC~\cite{shafiee2016isaac}) as our deployment prototype. Compared to traditional deployment, our proposed optimization framework achieves speed improvements of up to 60.43$\times$ and energy efficiency improvements of up to 1.89$\times$.







\section{\textbf{Motivation}}


\begin{figure}[!b]
    \vspace{-6mm}
  \setlength{\abovecaptionskip}{1.4mm}
  \setlength{\belowcaptionskip}{-0.cm}
  \centering
  \includegraphics{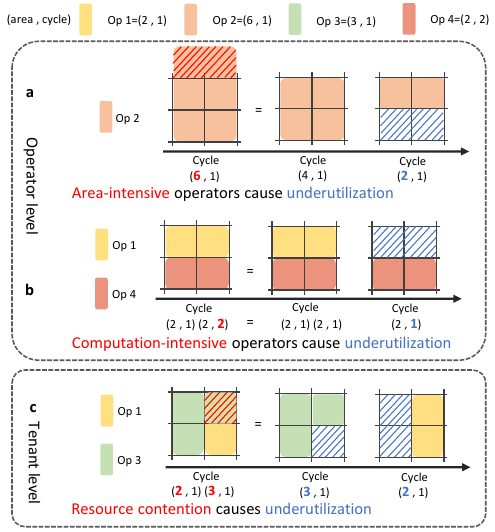}
  \caption{Challenges to deploy multi-tenant DNNs on ReRAM computing}
  \label{fig2}
\end{figure}

\subsection{\textbf{Resource Underutilization at the Operator Level}}

\begin{figure}[t]
  \setlength{\abovecaptionskip}{1.4mm}
  \setlength{\belowcaptionskip}{-0.cm}
  \centering
  \includegraphics{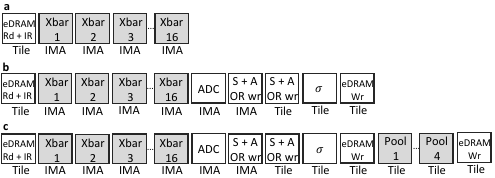}
  \caption{(a) Computational pipeline for writing data. (b) Pipelines for an operator without pooling layers. (c) Pipelines for an operator with subsequent pooling layers}
  \label{fig4}
  \vspace{-7mm}
\end{figure}

In ReRAM-based DNN computing, certain types of operators may have high computing area~(or cycles) demands but few cycles~(or area) requirements. This area-cycle imbalance may bottleneck the pipeline itself or block other parallel pipelines by occupying their computing area. This challenge is further amplified in a multi-tenant scenario, where accelerators are tasked with several simultaneous networks. These networks also construct their inter-layer computing pipelines. 
The resource underutilization introduced by computation-intensive or area-intensive operators becomes increasingly prominent in these highly parallelized multi-tenant DNNs.

Various operators need diverse computation times, among which computation-intensive operators may clog the pipeline and lead to resource underutilization. Unlike GPU-based deep learning frameworks where computation numbers across layers are relatively uniform, there can be up to millions of computations for computation-intensive layers and only one for other layers, leading to the 'Barrel Principle.' This principle implies that inefficiency in one stage limits overall system performance. As illustrated in Figure~\ref{fig2}b, deploying operator(2,1) and~(2,2) on a 4-unit area would waste 2 units per clock cycle.

At the operator level in multi-tenant DNNs, area-intensive operators usually have massive parameters and need large computing areas. Unlike convolutional layers requiring many computation clock cycles, a large on-chip area can update millions of parameters in one cycle. As Figure~\ref{fig2}a illustrates, using additional clock cycles when the on-chip computing area is insufficient may lead to underutilization. 
This over-occupation issue is pronounced in multi-tenant DNNs with smaller areas for each network. Though operator-splitting strategies have been proposed~\cite{7920854,tang2017binary}, integrating these into an overall optimization framework for complex multi-tenant DNNs remains a pressing challenge.

The multi-tenant deployment pattern makes the underutilization issue more severe than single networks. Simultaneously processing multi-tenant networks requires a larger deployment area and amplifies the latency due to computation-intensive operators, resulting in more write-and-compute cycles. This underutilization, stemming from extended computation cycles and excessive area requirement, not only hampers overall efficiency but also aggravates system latency, necessitating immediate attention.

\subsection{\textbf{Resource Competition at Tenant Level}}

A significant difference between ReRAM-based computing and conventional GPU/CPU is that computational resources are directly related to the allocated area in ReRAM platforms, which causes resource contention at the tenant level and extends the computation latency. At the tenant level in multi-tenant DNNs, improper area allocation can lead to resource contention and underutilization. When running multiple networks simultaneously on one ReRAM chip, overlapping re-write and processing regions can cause contention, as shown in Figure~\ref{fig2}c, where NN1 and NN2 compete for a unit area, leaving three units idle in two cycles. This leads to significant underutilization and reduced efficiency. The issue can be resolved by segregating parallel networks into separate areas to optimize computational time.

In a multi-tenant DNN scenario, the overall results consist of independent results from each tenant, which means the overall latency depends on the network with the highest processing time. Partitioning at the tenant level can mitigate resource contention but amplify the variances in processing time. Therefore, we need to partition computing resources according to the processing time of each tenant, which depends on the operator reconstruction at the operator level. In essence, whether at the operator level or the network level, optimizing in isolation may impact the optimization outcomes of the other level. This can result in a scenario where concentration shifts to local optimization rather than achieving global optimization. Hence, a joint optimization between operator level and tenant level is vital to schedule multi-tenant DNNs on ReRAM-based computing.

\section{Cross-Level Optimization Framework}

Figure~\ref{fig5} illustrate the proposed joint optimization framework, which include a profiler to analysis hardware computing performance, a intelligent and tenant aware hardware resource partition, and a fine-grained operators reconstruction.

\subsection{Classic Profiler and Inter-layer Parallelism}

After determining tenant architecture and accelerator topology, computation costs, latency, and energy are analyzed using the classic profiler, as described in~\cite{shafiee2016isaac}. With each clock cycle set to 100ns, the profiler calculates clock cycles for each operator on each tenant architecture. Figure~\ref{fig4} showcases this analysis, and processing time follows equation (1), with $C_{l}$ as clock cycles (100ns), $N_{b}$ as calculation bits, and $N_{p}$ as subsequent pooling cycles. Besides, for the energy calculation, consumption is determined by the power usage~(Table~\ref{tab:hardware}) of computational tiles~(according to tenant architecture and accelerator topology) and corresponding processing time~(as described above), while disregarding the static power consumption of idle components.

\vspace{-10pt}
\begin{equation}
T = C_{l} [(1+N_{b})+(6+N_{b}+N_{p}) ]
\end{equation}
\vspace{-10pt}

\begin{figure}[t]
  \setlength{\abovecaptionskip}{1.4mm}
  \setlength{\belowcaptionskip}{-0.cm}
  \centering
  \includegraphics{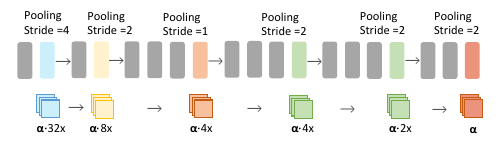}
  \caption{Classic inter-layer parallelism according to tenant architecture}
  \label{fig-3.1}
  \vspace{-5mm}
\end{figure}

\begin{figure*}[t]
  \setlength{\abovecaptionskip}{1.4mm}
  \setlength{\belowcaptionskip}{-0.cm}
  \centering
  \includegraphics{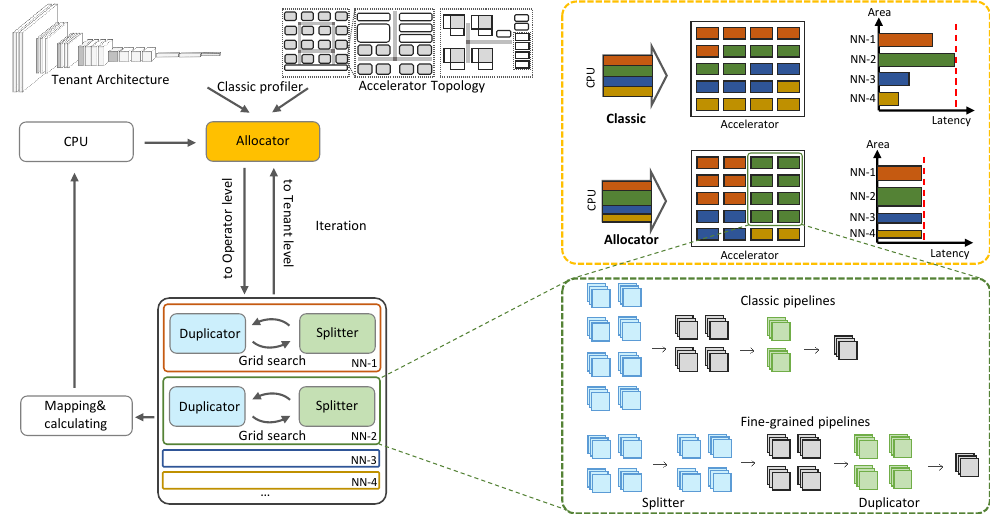}
  \caption{Overall optimization for ReRAM computing with multi-tenant DNNs deployment}
  \label{fig5}
  \vspace{-2mm}
\end{figure*}

The classic profiler considers not only the calculation pipeline within each operator but also between operators. As shown in Figure~\ref{fig-3.1}, traditional deployment augments inter-layer parallelism to create a high-throughput pipeline by duplicating convolutional kernels. When sequentially deployed, an operator with a stride of 2 needs two clock cycles for an output value. Networks like VGGs include multiple stride-$k$ operators, making deeper ones require $k$ exponential power cycles for one output. Analyzing the architecture and stride reveals a proportional relationship for kernel duplication.

\vspace{-8pt}
\begin{equation}
N_{p} = \prod_{i}^{n} s_{i}
\end{equation}
\vspace{-10pt}

The duplication count in inter-layer parallelism is determined using equation 2, where $i$ is the current layer, $n$ is the tenant's total layers, and $s_{i}$ is the stride of the next pooling layer. By combining Eq.1 and Eq.2, classic profilers can assess tenant computation time, accounting for inter-layer parallelism and the processing pipeline within an operator.

\renewcommand{\algorithmicrequire}{\textbf{Input:}}  
\renewcommand{\algorithmicensure}{\textbf{Output:}} 

\begin{algorithm}[htbp]
  \caption{A tenant-operator joint optimization algorithm for multi-tenant DNNs} 
  \label{alg::jointoptimization}
  \begin{algorithmic}[1]
    \Require
      $s_i$: network structure for tenant $i$;
      $s_c$: architecture of accelerator;
      $l$: acceptable delay between tenants;
    \Ensure
      deployment $d$
    \State initial $a_1$ = $a_2$ = $a_3$ = 1/3 computational area;

    \Function{allocateArea}{$s_1, s_2, s_3, s_c$}
    \While {$delay > l$}
        \State iterAllocation($a_1$,$a_2$,$a_3$)
        \State $t_i$ = processingTime($s_i$,$a_i$)
        \State delay = max($t_1$,$t_2$,$t_3$) - min($t_1$,$t_2$,$t_3$)
    \EndWhile
    \EndFunction
    \Function{processingTime}{$s_i$, $a_i$}
        \State gridSearch($\alpha$ , $\beta$)
        \State re-Op = duplicateandSplit($s_i$,$\alpha$,$\beta$)
        \State $time_{\alpha,\beta}$ = naiveDeploy(re-OP,$a_i$)
        \State save $time_{\alpha,\beta}$
        \State return min(time)
    \EndFunction
  \end{algorithmic}
\end{algorithm}
\vspace{-10pt}

\subsection{Intelligent Hardware Resource Partition}

Our main purpose at the tenant level is to divide separate regions for each tenant to avoid spatial resource contention and to intelligently partition computing area for each tenant that better suit their quantities of computation. The main steps are as follows. First, we divide the entire on-chip computing area into equal-sized regions for tenants. By doing so, although we have solved the spatial dependency, the same computing resource allocated to different tenants will cause different processing times and the tenant that completes the computation early will have a resource waste problem. Second, we adjust the computing resources by iteration, which can be controlled by learning rate $\eta$. After initializing the area of each region by dividing it equally, we apply grid search to find the best duplicate and split parameter for each tenant according to the computing area, tenant structure, and input size. The computation time for each tenant will be shown to us with fine-toned parameters. Then we switch a part of the area belonging to the fastest tenant to the slowest tenant according to the learning rate $\eta$ and find another pair of duplicate and split parameters in new iterations. Finally, the early-stopping part will intervene and stop the iteration once the area allocation this time is the same as two rounds ago. Our overall algorithm for multi-tenant DNNs for ReRAM-based computing is shown in Algorithm 1.

\subsection{Fine-grained Operators Reconstrcution}

 
In this section, we first introduce how to reconstruct imbalanced operators via \textbf{Duplicator} and \textbf{Splitter}. Then, we utilize the grid-search method to jointly optimize the duplicate and split parameters to obtain fine-grained reconstructed operators, referred to as re-operators. These re-operators meet the necessary criteria to compose our inter-layer pipeline, preventing underutilization caused by over-occupation and excessive computation.


\noindent \textbf{Duplicator:} In the classic deployment of single neural networks on ReRAM-based chips, researchers determine the duplication ratio based on the architecture of networks and further construct inter-layer pipelines. Our research aims to create the most computationally efficient fine-grained operators for a given configuration through joint duplication and splitting at the operator level. Rather than having individual duplication ratios for each layer, we introduce an overall duplicate parameter,$\alpha$, to adjust the duplication ratio while maintaining the original inter-layer ratio. Keeping this original inter-layer ratio has significant benefits. First, keeping the originally balanced parallelism can avoid pipeline stalls and ensure the pipeline's high throughput. Second, optimization of splitting and duplicating with separate replication ratios for each operator will lead to complex computation, especially in multi-tenant scenarios. Last, setting the duplicating ratio, $\alpha$, based on the traditional inter-layer pipeline guides the search space of $\alpha$ since our splitter will avoid duplicator to manually create new area-intensive operators.


\noindent \textbf{Splitter:} Operators like fc layers need substantial area for weight deployment but only brief computation time. Every two neurons between fc layers are connected with weights that accumulate into a large number. While convolutional operators require a few sliding filters for multiple computations. Once weights are deployed, results are achieved through a single calculation. Co-deploying fc with multi-computation operators such as conv leads to idleness and resource underutilization.

To reconstruct fine-grained operators and enhance computational efficiency at the tenant level, our proposed framework does splitting and duplicating simultaneously. As we mentioned before, the number of rows to be occupied in crossbar arrays is determined by the dimension of the input feature map. Only split into integer multiples of crossbar array rows can maximize the utilization of computing areas. To control the size of the split, we set a hyper-parameter $\beta$. 
According to our hyper-parameter $\beta$ that controls the size of the split, area-intensive operators that need to be deployed larger than $\beta$ will be split into smaller ones. When beta is relatively large, the original structure of the network can be better preserved. But it may also not be able to utilize the full computational area because of the coarse granularity. When beta is small, the whole network will be split into small pieces, and easy to fill the whole calculating area. In that way, the original area-intensive operators may be transferred into computation-intensive operators and also hurt the overall latency.

Since these parameters are interconnected, we employ the grid search method to identify the optimal and harmonious values. The joint optimization of these two hyperparameters can balance the duplication rate and splitting rate, ensuring the throughput while making sure that each operator neither requires excessive computations nor over-occupation of certain areas. By joint splitting and duplicating, we can build fine-grained, high-throughput processing pipelines that fill the given calculation area at minimal waste, while completing their own calculation in a similar time to prepare for the next round of writing weights.

\section{\textbf{Experiments}}
\subsection{Experiments setup}

\begin{table*}[t]
\caption{Neural networks of different complexity that are used to in multi-tenant DNNs}
\begin{tabular}{|ccc|ccccc|}
\hline
\multicolumn{1}{|c|}{\textbf{DNN1}} &
  \multicolumn{1}{c|}{\textbf{DNN2}} &
  \textbf{DNN3} &
  \multicolumn{1}{c|}{\textbf{DNN4}} &
  \multicolumn{1}{c|}{\textbf{VGG11}} &
  \multicolumn{1}{c|}{\textbf{VGG13}} &
  \multicolumn{1}{c|}{\textbf{VGG16}} &
  \textbf{VGG19} \\ \hline
\multicolumn{1}{|c|}{3x3,64(1)} &
  \multicolumn{1}{c|}{7x7,96(1)} &
  7x7,16(1) &
  \multicolumn{1}{c|}{7x7,16(1)} &
  \multicolumn{1}{c|}{3x3,64(1)} &
  \multicolumn{1}{c|}{3x3,64(2)} &
  \multicolumn{1}{c|}{3x3,64(2)} &
  3x3,64(2) \\ \hline
\multicolumn{1}{|c|}{2x2,maxpool/2} &
  \multicolumn{1}{c|}{2x2,maxpool/2} &
  2x2,maxpool/2 &
  \multicolumn{1}{c|}{2x2,maxpool/2} &
  \multicolumn{1}{c|}{2x2,maxpool/2} &
  \multicolumn{1}{c|}{2x2,maxpool/2} &
  \multicolumn{1}{c|}{2x2,maxpool/2} &
  2x2,maxpool/2 \\ \hline
\multicolumn{1}{|c|}{3x3,128(1)} &
  \multicolumn{1}{c|}{3x3,256(1)} &
  3x3,48(2) &
  \multicolumn{1}{c|}{3x3,48(5)} &
  \multicolumn{1}{c|}{3x3,128(1)} &
  \multicolumn{1}{c|}{3x3,128(2)} &
  \multicolumn{1}{c|}{3x3,128(2)} &
  3x3,128(2) \\ \hline
\multicolumn{1}{|c|}{2x2,maxpool/2} &
  \multicolumn{1}{c|}{2x2,maxpool/2} &
  2x2,maxpool/2 &
  \multicolumn{1}{c|}{2x2,maxpool/2} &
  \multicolumn{1}{c|}{2x2,maxpool/2} &
  \multicolumn{1}{c|}{2x2,maxpool/2} &
  \multicolumn{1}{c|}{2x2,maxpool/2} &
  2x2,maxpool/2 \\ \hline
\multicolumn{1}{|c|}{3x3,256(2)} &
  \multicolumn{1}{c|}{3x3,512(1)} &
  3x3,64(2) &
  \multicolumn{1}{c|}{3x3,64(5)} &
  \multicolumn{1}{c|}{3x3,256(2)} &
  \multicolumn{1}{c|}{3x3,256(2)} &
  \multicolumn{1}{c|}{3x3,256(3)} &
  3x3,256(4) \\ \hline
\multicolumn{1}{|c|}{2x2,maxpool/2} &
  \multicolumn{1}{c|}{2x2,maxpool/2} &
  2x2,maxpool/2 &
  \multicolumn{1}{c|}{2x2,maxpool/2} &
  \multicolumn{1}{c|}{2x2,maxpool/2} &
  \multicolumn{1}{c|}{2x2,maxpool/2} &
  \multicolumn{1}{c|}{2x2,maxpool/2} &
  2x2,maxpool/2 \\ \hline
\multicolumn{1}{|c|}{3x3,512(2)} &
  \multicolumn{1}{c|}{3x3,512(1)} &
   &
  \multicolumn{1}{c|}{3x3,64(5)} &
  \multicolumn{1}{c|}{3x3,512(2)} &
  \multicolumn{1}{c|}{3x3,512(2)} &
  \multicolumn{1}{c|}{\begin{tabular}[c]{@{}c@{}}3x3,512(2)\\ 3x3,256(1)\end{tabular}} &
  3x3,512(4) \\ \hline
\multicolumn{1}{|c|}{2x2,maxpool/2} &
  \multicolumn{1}{c|}{2x2,maxpool/2} &
   &
  \multicolumn{1}{c|}{2x2,maxpool/2} &
  \multicolumn{1}{c|}{2x2,maxpool/2} &
  \multicolumn{1}{c|}{2x2,maxpool/2} &
  \multicolumn{1}{c|}{2x2,maxpool/2} &
  2x2,maxpool/2 \\ \hline
\multicolumn{1}{|c|}{3x3,512(2)} &
  \multicolumn{1}{c|}{} &
   &
  \multicolumn{1}{c|}{} &
  \multicolumn{1}{c|}{3x3,512(2)} &
  \multicolumn{1}{c|}{3x3,512(2)} &
  \multicolumn{1}{c|}{\begin{tabular}[c]{@{}c@{}}3x3,512(2)\\ 3x3,256(1)\end{tabular}} &
  3x3,512(4) \\ \hline
\multicolumn{1}{|c|}{2x2,maxpool/2} &
  \multicolumn{1}{c|}{} &
   &
  \multicolumn{1}{c|}{} &
  \multicolumn{1}{c|}{2x2,maxpool/2} &
  \multicolumn{1}{c|}{2x2,maxpool/2} &
  \multicolumn{1}{c|}{2x2,maxpool/2} &
  2x2,maxpool/2 \\ \hline
\multicolumn{3}{|c|}{fc-512(2)} &
  \multicolumn{5}{c|}{fc-4096(2)} \\ \hline
\multicolumn{3}{|c|}{fc-100(1)} &
  \multicolumn{5}{c|}{fc-1000(1)} \\ \hline
\end{tabular}
\label{tab:setup}
\vspace{-4mm}
\end{table*}

\begin{table}[htbp]
\caption{Hardware parameters from ISAAC~\cite{shafiee2016isaac}}
\centering
\begin{tabular}{|llll|}
\hline
\multicolumn{4}{|c|}{ISAAC Tile at 1.2GHz, 0.37mm2} \\ \hline
\multicolumn{1}{|l|}{Component} &
  \multicolumn{1}{l|}{Params} &
  \multicolumn{1}{l|}{Spec} &
  Power \\ \hline
\multicolumn{1}{|l|}{\begin{tabular}[c]{@{}l@{}}eDRAM\\ Buffer\end{tabular}} &
  \multicolumn{1}{l|}{\begin{tabular}[c]{@{}l@{}}size\\ num\_banks\\ bus\_width\end{tabular}} &
  \multicolumn{1}{l|}{\begin{tabular}[c]{@{}l@{}}64 KB\\ 4\\ 256 b\end{tabular}} &
  20.7 mW \\ \hline
\multicolumn{1}{|l|}{\begin{tabular}[c]{@{}l@{}}eDRAM\\ -to-IMA bus\end{tabular}} &
  \multicolumn{1}{l|}{num\_wire} &
  \multicolumn{1}{l|}{384} &
  7 mW \\ \hline
\multicolumn{1}{|l|}{Router} &
  \multicolumn{1}{l|}{\begin{tabular}[c]{@{}l@{}}flit\_size\\ num\_port\end{tabular}} &
  \multicolumn{1}{l|}{\begin{tabular}[c]{@{}l@{}}32\\ 8\end{tabular}} &
  42 mW \\ \hline
\multicolumn{1}{|l|}{\begin{tabular}[c]{@{}l@{}}Sigmoid\\ S+A\\ MaxPool\\ OR\end{tabular}} &
  \multicolumn{1}{l|}{\begin{tabular}[c]{@{}l@{}}number\\ number\\ number\\ size\end{tabular}} &
  \multicolumn{1}{l|}{\begin{tabular}[c]{@{}l@{}}2\\ 1\\ 1\\ 3 KB\end{tabular}} &
  \begin{tabular}[c]{@{}l@{}}0.53 mW\\ 0.05 mW\\ 0.4 mW\\ 1.68 mW\end{tabular} \\ \hline
\multicolumn{1}{|l|}{\textbf{Total}} &
  \multicolumn{1}{l|}{} &
  \multicolumn{1}{l|}{} &
  \textbf{40.9}  mW \\ \hline
\multicolumn{4}{|c|}{IMA properties (12 IMA per tile)} \\ \hline
\multicolumn{1}{|l|}{ADC} &
  \multicolumn{1}{l|}{\begin{tabular}[c]{@{}l@{}}resolution\\ frequency\\ number\end{tabular}} &
  \multicolumn{1}{l|}{\begin{tabular}[c]{@{}l@{}}8 bits\\ 1.2 GSps\\ 8\end{tabular}} &
  16 mW \\ \hline
\multicolumn{1}{|l|}{DAC} &
  \multicolumn{1}{l|}{\begin{tabular}[c]{@{}l@{}}resolution\\ number\end{tabular}} &
  \multicolumn{1}{l|}{\begin{tabular}[c]{@{}l@{}}1 bits\\ 8 $\times$ 128\end{tabular}} &
  4 mW \\ \hline
\multicolumn{1}{|l|}{S+H} &
  \multicolumn{1}{l|}{number} &
  \multicolumn{1}{l|}{8 $\times$ 128} &
  10 $\mu$W \\ \hline
\multicolumn{1}{|l|}{\begin{tabular}[c]{@{}l@{}}Memristor\\ array\end{tabular}} &
  \multicolumn{1}{l|}{\begin{tabular}[c]{@{}l@{}}number\\ size\\ bits per cell\end{tabular}} &
  \multicolumn{1}{l|}{\begin{tabular}[c]{@{}l@{}}8\\ 128 $\times$ 128\\ 2\end{tabular}} &
  2.4 mW \\ \hline
\multicolumn{1}{|l|}{\begin{tabular}[c]{@{}l@{}}S+A\\ IR\\ OR\end{tabular}} &
  \multicolumn{1}{l|}{\begin{tabular}[c]{@{}l@{}}number\\ size\\ size\end{tabular}} &
  \multicolumn{1}{l|}{\begin{tabular}[c]{@{}l@{}}4\\ 2KB\\ 256B\end{tabular}} &
  \begin{tabular}[c]{@{}l@{}}0.2 mW\\ 1.24 mW\\ 0.23 mW\end{tabular} \\ \hline
\multicolumn{1}{|l|}{\textbf{IMA Total}} &
  \multicolumn{1}{l|}{number} &
  \multicolumn{1}{l|}{12} &
  \textbf{289} mW \\ \hline
\multicolumn{1}{|l|}{\begin{tabular}[c]{@{}l@{}}\textbf{1 Tile Total}\\ \textbf{168 Tile Total}\end{tabular}} &
  \multicolumn{1}{l|}{} &
  \multicolumn{1}{l|}{} &
  \begin{tabular}[c]{@{}l@{}}\textbf{330} mW\\ \textbf{55.4} W\end{tabular} \\ \hline
\multicolumn{1}{|l|}{\textbf{Chip Total}} &
  \multicolumn{1}{l|}{} &
  \multicolumn{1}{l|}{} &
  \textbf{65.8} W \\ \hline
\end{tabular}
\label{tab:hardware}
\end{table}

\noindent \textbf{Various sizes of Multi-tenant DNNs:}
As shown in Table~\ref{tab:setup}, we used eight different base networks which include both shallow networks and complex networks to evaluate the speedup and energy of our deployment method for different complexity of multi-tenant DNNs.

\noindent \textbf{Comparation baseline:} The joint optimization method presented in this paper does not rely on accelerators with a specific topology. Instead, it iteratively partitions at the tenant level and reconstructs fine-grained pipelines at the operator level, adapting to various accelerator architectures. Therefore, a simple and classic accelerator can minimize unrelated factors and focus on the optimization framework itself. We did all the experiments and analysis on ISAAC~\cite{shafiee2016isaac}, using this classic accelerator to explain the operational principles, observe the acceleration effects, and provide insights for future researchers.

\noindent \textbf{Inference setup:} With the power/area value from~\cite{shafiee2016isaac} as is shown in Table~\ref{tab:hardware}, we first verify the overall speedup of our joint optimization framework. Then we show the effectiveness of optimization at the tenant level and operator level respectively. We set eight distinct multi-tenant DNNs on three chip sizes to validate the optimization performance of our framework within the context of ReRAM-based computation. These computations span various complexities of multi-tenant DNNs and different accelerator topologies.

\subsection{Energy Consumption Analyze}

After using our joint optimization for deployment, we follow the original principle of accelerators to calculate energy as described in sec 3A. Static energy consumption is ignored for the idle part.

Table \ref{tab:results} shows that in small-sized chips (chip1), 7 of 8 multi-tenant DNNs reduced energy use, owing to resource contention. Coarse-grained operators in a tenant hinder full utilization of on-chip resources, with smaller chips amplifying tenant-level contention and operator-level underutilization, thus increasing energy consumption. Our framework targets these issues by eliminating contentions at the tenant level and reconstructing operators to enhance the efficiency of on-chip resource utilization, these optimization reduces energy consumption. For medium and large-sized chips, a balance must be struck between underutilization from resource contention and accumulated resource waste from dividing into multiple independent regions. Allocating distinct computation areas for each tenant improves utilization but fails to reach 100\% efficiency, leading to an increased waste of area with more tenants. Our framework can cut energy usage in medium-sized chip experiments (chip2), though less effectively than in smaller chips (chip1). In larger chips (chip3) with less contention, our framework doesn't yield energy savings.

\subsection{Speed-up Evaluation and Analysis}

\begin{table}[t]
\begin{threeparttable}
\caption{Summary of experiments results}
\centering
\captionsetup{singlelinecheck = false, justification=justified, font=footnotesize}
\tabcolsep 2.6pt
\begin{tabular}{lcccccc}

\hline
{\color[HTML]{CB0000} }                 
& \tabincell{c}{Multi-tenant\\ DNN} & \tabincell{c}{origin \\energy(mJ)}  & \tabincell{c}{optimi \\energy(mJ)}
  & \tabincell{c}{overall \\speedup} & \tabincell{c}{te-level \\speedup} & \tabincell{c}{op-level \\speedup} \\ \hline
   & MT1   &19.75 &10.44    & 4.48  & 1.12   & 3.99  \\
   & MT2  &37.49 &27.07    & 1.79  & 1.08   & 1.66  \\
   & MT3  &33.76  &23.02    & 1.85  & 1.13   & 1.64  \\
   & MT4     &17.43  &10.70    & 12.6  & 1.03 & 12.23  \\
   & MT5     &11.57  &12.09    & 22.53 & 1.06    & 21.25  \\
   & MT6 &36.50  &20.58 & 4.8   & 1.15   & 4.17  \\
   & MT7 &46.42  &35.45 & 1.75  & 1.26   & 1.39  \\
\multirow{-8}{*}{\rotatebox{90}{\tabincell{c}{chip1\\(168-12-8-128)}}}     & MT8    &37.95  &27.09           & 1.96                                    & 1.22                    & 1.61                \\ \hline
   & MT1   &19.69  &15.38    & 5.92  & 1.06    & 5.58  \\
   & MT2  &37.42  &38.83    & 1.98  & 0.87  & 2.28 \\
   & MT3  &33.69  &32.33    & 2.17  & 0.97   & 2.24 \\
   & MT4     &17.40  &13.19    & 16.15 & 1.02 & 15.83  \\
   & MT5     &11.56  &12.09    & 30.85 & 1.01    & 30.84 \\
   & MT6 &36.50  &31.50 & 7.2   & 1.11   & 6.49  \\
   & MT7 &46.42  &46.40   & 2.01  & 1.05    & 1.91  \\
\multirow{-8}{*}{\rotatebox{90}{\tabincell{c}{chip2\\(256-12-12-128)}}}    & MT8    &37.95  &35.44           & 2.32                                & 1.07                & 2.17            \\ \hline
\end{tabular}
\begin{tablenotes}
\footnotesize
\item[] MT1: DNN4+VGG11+VGG16 \quad
        MT2: VGG13+VGG16+VGG19
\item[] MT3: VGG11+VGG13+VGG19 \quad
        MT4: DNN1+DNN2+DNN3
\item[] MT5: DNN2+DNN3+DNN4 \quad
        MT6: DNN1+DNN3+VGG16+VGG19
\item[] MT7: VG11+VGG13+VGG16+VGG19 \\
        MT8: DNN1+VGG11+VGG13+VGG16
\end{tablenotes}
\label{tab:results}
\vspace{-6mm}
\end{threeparttable}
\end{table}

Following the processing principles of original accelerators~(section 2A), we get the result Table~\ref{tab:results}. After normalizing ISAAC baselines into 1, our latency speedup varies between 1.78 and 60.43 across all experiments, which indicates that our cross-level framework can enhance the efficiency of on-chip resources. For a detailed analysis, we have two observations. 
First, multi-tenant DNNs composed of shallow networks (such as MT4 and MT5) demonstrate a better speedup on various sizes of ReRAM-based accelerators, with values ranging from 12.60 to 60.45. This might be because complex multi-tenant DNNs have a wider array of operators available for selection during deployment while shallow networks present fewer candidate operators for selection and depend more heavily on external tools like ours to alleviate this temporal-spatial imbalance.
Second, four out of the five multi-tenant DNNs demonstrate superior speedup on larger-sized accelerators, which indicates that a larger-sized accelerator might enhance the efficiency of computational area usage during optimization. Fine-grained re-operators strive to maximize the use of the available computational area on the ReRAM-based accelerators. But some computational areas will inevitably remain idled. The percentage of wasted area relative to the total computing area diminishes on a larger accelerator, which means the larger the computational area of an accelerator, the greater the efficiency of computational resource usage.

Following the discussion on the overall speedup produced by our cross-level optimization, we delve deeper into the separate contributions made to computational efficiency by iterative partitioning at the tenant level and fine-grained reconstruction at the operator level, facilitated through ablation analysis. As illustrated in table~\ref{tab:results}, we fixed the cross-level experimental speedup at 100\%, analyzing the relative speedup from optimizations at both tenant and operator levels respectively.

From the view of chip topology, iterative allocation at the tenant level proves more effective in experiments conducted on smaller-sized chips (chip1), achieving 14\% to 24\% of the overall contribution.
For experiments conducted on medium-sized accelerators~(chip2) and large-sized accelerators~(chip3), most of the speedup is attributed to operator-level reconstruction. In conclusion, iterative region allocation at the tenant level performs better on chips with relatively smaller topological sizes, while fine-grained operator reconstruction contributes significantly to latency reduction. This might be because identical multi-tenant DNNs running on smaller topology chips encounter more competition between tenants, leading to greater resource underutilization due to these contentions.

\section{Conclusion and Discussion}

In this work, we proposed a cross-level ReRAM-based in-memory framework for multi-tenant DNNs. In our joint optimization framework, we initially allocate different computational areas iteratively for different tenants at the tenant level. Then, for each individual tenant based on the assigned computational resources, we perform a fine-grained operator reconstruction method which consists of splitters and duplicators to rebuild processing pipelines. In this way, we achieved a 1.75x acceleration and energy reduction in various topological structures of chips and different multi-tenant DNNs.

\section{Acknowlegement}

\bibliographystyle{_sty/ieee}
\bibliography{_bib/ASPDAC-2023}

\begin{thebibliography}{10}
\providecommand{\url}[1]{#1}
\csname url@rmstyle\endcsname
\providecommand{\newblock}{\relax}
\providecommand{\bibinfo}[2]{#2}
\providecommand\BIBentrySTDinterwordspacing{\spaceskip=0pt\relax}
\providecommand\BIBentryALTinterwordstretchfactor{4}
\providecommand\BIBentryALTinterwordspacing{\spaceskip=\fontdimen2\font plus
\BIBentryALTinterwordstretchfactor\fontdimen3\font minus
  \fontdimen4\font\relax}
\providecommand\BIBforeignlanguage[2]{{%
\expandafter\ifx\csname l@#1\endcsname\relax
\typeout{** WARNING: IEEEtran.bst: No hyphenation pattern has been}%
\typeout{** loaded for the language `#1'. Using the pattern for}%
\typeout{** the default language instead.}%
\else
\language=\csname l@#1\endcsname
\fi
#2}}

\bibitem{chen2016eyeriss}
Y.-H. Chen, \emph{et~al.}, ``Eyeriss: An energy-efficient reconfigurable
  accelerator for deep convolutional neural networks,'' \emph{IEEE journal of
  solid-state circuits}, vol.~52, no.~1, pp. 127--138, 2016.

\bibitem{chen2018regan}
F.~Chen, \emph{et~al.}, ``Regan: A pipelined reram-based accelerator for
  generative adversarial networks,'' in \emph{2018 23rd Asia and South Pacific
  Design Automation Conference (ASP-DAC)}.\hskip 1em plus 0.5em minus
  0.4em\relax IEEE, 2018, pp. 178--183.

\bibitem{zheng2020lattice}
Q.~Zheng, \emph{et~al.}, ``Lattice: An adc/dac-less reram-based
  processing-in-memory architecture for accelerating deep convolution neural
  networks,'' in \emph{2020 57th ACM/IEEE Design Automation Conference
  (DAC)}.\hskip 1em plus 0.5em minus 0.4em\relax IEEE, 2020, pp. 1--6.

\bibitem{mao2018versatile}
M.~Mao, \emph{et~al.}, ``A versatile reram-based accelerator for convolutional
  neural networks,'' in \emph{2018 IEEE International Workshop on Signal
  Processing Systems (SiPS)}.\hskip 1em plus 0.5em minus 0.4em\relax IEEE,
  2018, pp. 211--216.

\bibitem{jeon2019analysis}
M.~Jeon, \emph{et~al.}, ``Analysis of
  $\{$Large-Scale$\}$$\{$Multi-Tenant$\}$$\{$GPU$\}$ clusters for $\{$DNN$\}$
  training workloads,'' in \emph{2019 USENIX Annual Technical Conference
  (USENIX ATC 19)}, 2019, pp. 947--960.

\bibitem{yu2022survey}
F.~Yu, \emph{et~al.}, ``A survey of multi-tenant deep learning inference on
  gpu,'' \emph{arXiv preprint arXiv:2203.09040}, 2022.

\bibitem{huang2021close}
Y.~Huang, \emph{et~al.}, ``A close look at multi-tenant parallel cnn inference
  for autonomous driving,'' in \emph{Network and Parallel Computing: 17th IFIP
  WG 10.3 International Conference, NPC 2020, Zhengzhou, China, September
  28--30, 2020, Revised Selected Papers}.\hskip 1em plus 0.5em minus
  0.4em\relax Springer, 2021, pp. 92--104.

\bibitem{jeon2016multi}
S.~Jeon \emph{et~al.}, ``Multi-robot multi-task allocation for hospital
  logistics,'' in \emph{2016 18th International Conference on Advanced
  Communication Technology (ICACT)}.\hskip 1em plus 0.5em minus 0.4em\relax
  IEEE, 2016, pp. 339--341.

\bibitem{chen2018emat}
F.~Chen \emph{et~al.}, ``Emat: an efficient multi-task architecture for
  transfer learning using reram,'' in \emph{2018 IEEE/ACM International
  Conference on Computer-Aided Design (ICCAD)}.\hskip 1em plus 0.5em minus
  0.4em\relax ACM, 2018, pp. 1--6.

\bibitem{tambe2021edgebert}
T.~Tambe, \emph{et~al.}, ``Edgebert: Sentence-level energy optimizations for
  latency-aware multi-task nlp inference,'' in \emph{MICRO-54: 54th Annual
  IEEE/ACM International Symposium on Microarchitecture}, 2021, pp. 830--844.

\bibitem{ankit2020panther}
A.~Ankit, \emph{et~al.}, ``Panther: A programmable architecture for neural
  network training harnessing energy-efficient reram,'' \emph{IEEE Transactions
  on Computers}, vol.~69, no.~8, pp. 1128--1142, 2020.

\bibitem{shafiee2016isaac}
A.~Shafiee, \emph{et~al.}, ``Isaac: A convolutional neural network accelerator
  with in-situ analog arithmetic in crossbars,'' \emph{ACM SIGARCH Computer
  Architecture News}, vol.~44, no.~3, pp. 14--26, 2016.

\bibitem{7920854}
L.~Song, \emph{et~al.}, ``Pipelayer: A pipelined reram-based accelerator for
  deep learning,'' in \emph{2017 IEEE International Symposium on High
  Performance Computer Architecture (HPCA)}, 2017, pp. 541--552.

\bibitem{tang2017binary}
T.~Tang, \emph{et~al.}, ``Binary convolutional neural network on rram,'' in
  \emph{2017 22nd Asia and South Pacific Design Automation Conference
  (ASP-DAC)}.\hskip 1em plus 0.5em minus 0.4em\relax IEEE, 2017, pp. 782--787.

\end{thebibliography}

\end{document}